\documentclass[english,aps,prl,twocolumn,floats,showpacs]{revtex4}
\usepackage[T1]{fontenc}
\usepackage[latin9]{inputenc}
\usepackage{amsmath}
\usepackage{amssymb}
\usepackage{graphicx}
\usepackage{esint}

\makeatletter

\@ifundefined{textcolor}{}
{%
 \definecolor{BLACK}{gray}{0}
 \definecolor{WHITE}{gray}{1}
 \definecolor{RED}{rgb}{1,0,0}
 \definecolor{GREEN}{rgb}{0,1,0}
 \definecolor{BLUE}{rgb}{0,0,1}
 \definecolor{CYAN}{cmyk}{1,0,0,0}
 \definecolor{MAGENTA}{cmyk}{0,1,0,0}
 \definecolor{YELLOW}{cmyk}{0,0,1,0}
}

\usepackage{babel}
\usepackage{bm}

\makeatother

\usepackage{babel}

\begin{document}

\title{Topological order generated by random field in a 2D exchange model}

\author{E. M. Chudnovsky and D. A. Garanin}

\affiliation{Physics Department, Herbert H. Lehman College and Graduate School, The City University of New York \\
 250 Bedford Park Boulevard West, Bronx, New York 10468-1589, USA}

\date{\today}
\begin{abstract}
We study a 2D exchange model with a weak static random field on lattices containing over one hundred million spins. Ferromagnetic correlations persist on the Imry-Ma scale inversely proportional to the random-field strength and decay exponentially at greater distances. We find that the average energy of the correlated area is close to the ground-state energy of a skyrmion, while the topological charge of the area is close to $\pm 1$. Correlation function of the topological charge density exhibits oscillations with a period determined by the ferromagnetic correlation length, while its Fourier transform exhibits a maximum. These findings suggest that static randomness transforms a 2D ferromagnetic state into a skyrmion-antiskyrmion glass. 
\end{abstract}

\pacs{75.50.Lk,12.39.Dc,05.10.-a}

\maketitle

Studies of static randomness in field-theory models have a long history. They apply to amorphous magnets and spin glasses \cite{CSS,Sellmyer-PRL1986,Tejada-PRB91,3D-DG}, flux lattices in superconductors \cite{Larkin-JETP1970,Blatter-RMP1994,Gingras-Huse-PRB1996,Nattermann-2000,SI}, magnetic bubble and skyrmion lattices \cite{bubbles,Stosic,Zeissler}, charge density waves \cite{Efetov-77,Lee,Gruner-RMP88,Millis-PRL15,Tom-EC-PRB2015}, liquid crystals and polymer physics \cite{LC,Radzihovsky}, and He-3 in aerogel \cite{Volovik,Pollanen-NatPhys13}. It has been long understood that the effect of a static random field (RF) on the long-range order is stronger than the effect of thermal fluctuations. In 1975 Imry and Ma (IM) made a general observation \cite{ImryMa} that static randomness, no matter how week, destroys the long-range order in less than $d = 4$ dimensions in systems with continuous-symmetry order parameter. According to the IM argument the correlated area scales as $H_R^{2/(d-4)}$ with the strength $H_R$ of the RF. Such correlated regions received the name of IM domains. While this concept was widely used by the experimentalists in application to various physical systems it was later questioned by theorists \cite{Cardy-PRB1982,Villain-ZPB1984,Nattermann,Korshunov-PRB1993,Giamarchi,Orland,Kierfield,Feldman,Bogner,LeDoussal} who applied the renormalization group, variational and replica-symmetry breaking methods to the problem. They argued that static randomness must lead to a defect-free Bragg glass characterized by only a power-law decay of correlations. More recent large-scale numerical simulations of RF systems, accompanied by analytical work \cite{PGC-PRL2014}, have shown that exponential decay of correlations does occur in the absence of topological defects. In, e.g., spin systems with $n$ spin components this requires $n > d + 1$. All problems of practical interest, however, correspond to $n \leq d + 1$, when topological defects are present. For such problems the dispute about the nature of the glass state created by static randomness has never been settled. 

In this Letter we study the borderline case, $n = d + 1$, of a three-component spin field in two dimensions. It possesses nonsingular topological objects, skyrmions \cite{SkyrmePRC58}, as compared to singular objects for $ n < d + 1$ (e.g., vortices in 2D and 3D XY models). The absence of the Bragg glass in two dimensions was first noticed by Daniel Fisher et al. \cite{Fisher-PRL99} who argued that a pinned elastic medium would be unstable to dislocations. A similar argument exists for a 2D ferromagnet. The scale invariance of the pure continuous exchange model in two dimensions makes the ground-state energy of the skyrmion, $4\pi J$ (with $J$ being the exchange constant), independent of the skyrmion size $\lambda$. In a crystal lattice, violation of the scale invariance by the finite atomic spacing, $a$, adds the term proportional to $-J(a/\lambda)^2$ to the energy of the skyrmion, forcing it to collapse \cite{CCG-PRB2012}. This changes in the presence of the RF. Fluctuations of the RF make the  energy of its interaction with the skyrmion scale as $-H_R (\lambda/a)$ \cite{CG-NJP}, thus forcing sufficiently large skyrmions to blow up rather than collapse. As we shall see, however, in the absence of the external field, the ferromagnetic order that is needed for the skyrmions to exist, in accordance with the IM argument is limited to areas of size $R_f \propto 1/H_R$. It is therefore plausible that IM domains in a 2D RF system are made by skyrmions and antiskyrmions of average size $\lambda \sim R_f$. In what follows we will provide quantitative support to this picture by studying topological structure of the disordered state on lattices containing over $10^8$ spins.

The model is described by the Hamiltonian
\begin{eqnarray}\label{model}
H & = & -\frac{J}{2}\sum_{<ij>} {\bm \sigma}_i\cdot{\bm \sigma}_j -\sum_i {\bm \sigma}_i \cdot{\bf H}_{Ri} \\
& = & \int d^2 r \left[ \frac{\alpha}{2}\left(\frac{\partial
S_{b}}{\partial r_{\beta}}\right)^2 - {\bf S}\cdot {\bf H}_R \right].
\end{eqnarray}
The first formula corresponds to the discrete lattice version of the model, with ${\bm \sigma}_i$ being the spin at the $i$-th lattice site, and $<ij>$ meaning summation over the nearest neighbors. The second formula provides the continuous field-theory counterpart of the model with $\alpha$ being the exchange stiffness, index $b = 1,2,3$ indicating the components of the spin field ${\bf S}(x,y)$ of constant length $S_0$, and index $\beta = x,y$ indicating the components of the radius-vector in the $xy$ plane. The discrete and continuous models are related according to  $\sum_i = \int {d^2 r}/{a^2}$, ${\bm \sigma}_i = a^2{\bf S}({\bf r}_i)$, $\alpha = Ja^4$.

Mapping of the unit sphere represented by ${\bf s}(x,y) = {\bf S}(x,y)/S_0$ onto the $xy$ coordinate plane generates classes of homotopy \cite{BelPolJETP75} that describe skyrmions and antiskyrmions of quantized topological charge $Q = 0, \pm 1, \pm 2 , ...$, given by
\begin{equation}\label{Q}
Q = \int \frac{d^2 r}{8\pi} \epsilon_{\alpha\beta} s_a \epsilon_{abc} \frac{\partial s_b}{\partial r_\alpha}\frac{\partial s_c}{\partial r_\beta} = \int \frac{dx dy}{4\pi} \: {\bf s}\cdot \frac{\partial {\bf s}}{\partial x} \times\frac{\partial {\bf s}}{\partial y}.
\end{equation}
The quantity $q({\bf r}) = \frac{1}{4\pi}\, {\bf s}\cdot \left(\frac{\partial {\bf s}}{\partial x} \times\frac{\partial {\bf s}}{\partial y}\right)$ under the integral has the meaning of the topological charge density (TCD). We are interested in the spin-spin correlation function (CF), $\langle {\bf s}({\bf r}_1) \cdot {\bf s}({\bf r}_2)\rangle$, and the CF of the TCD, $\langle q({\bf r}_1) q({\bf r}_2)\rangle$. The first has been intensively studied for RF systems in the past while the second received little attention. As we shall see it sheds a new light on the structure of the disordered state. 

In the numerical work we use periodic boundary conditions and a collinear initial condition (CIC) for the spins. The latter corresponds to all spins initially aligned in one direction, which would be the ground state in the absence of the RF. The system prepared with the CIC is allowed to evolve to a minimum energy state in the presence of the RF  which direction is chosen randomly at each lattice site. Our numerical method searches for the energy minimum by combining sequential rotations of the spins towards the direction of the local effective field, ${\bf H}_{i,{\rm eff}}=-\delta H/\delta {\bm \sigma}_i$, with the energy-conserving spin flips, ${\bm \sigma}_{i}\rightarrow2({\bm \sigma}_{i}\cdot{\bf H}_{i,{\rm eff}}){\bf H}_{i,{\rm eff}}/H_{i,{\rm eff}}^{2}-{\bm \sigma}_{i}$. The two are applied with probabilities $\gamma$ and $1-\gamma$ respectively; $\gamma$ playing the role of the relaxation constant. The method has high efficiency for glassy systems under the condition $\gamma \ll1$ \cite{PGC-PRL2014}. The largest-scale computation has been done on a square lattice containing $10240 \times 10240$ spins. In numerical work we used $J = 1$ and $|{\bm \sigma}_i| = 1$, with all results easily rescaled for arbitrary $J$ and $|{\bm \sigma}_i|$. 

\begin{figure}
\includegraphics[width=90mm]{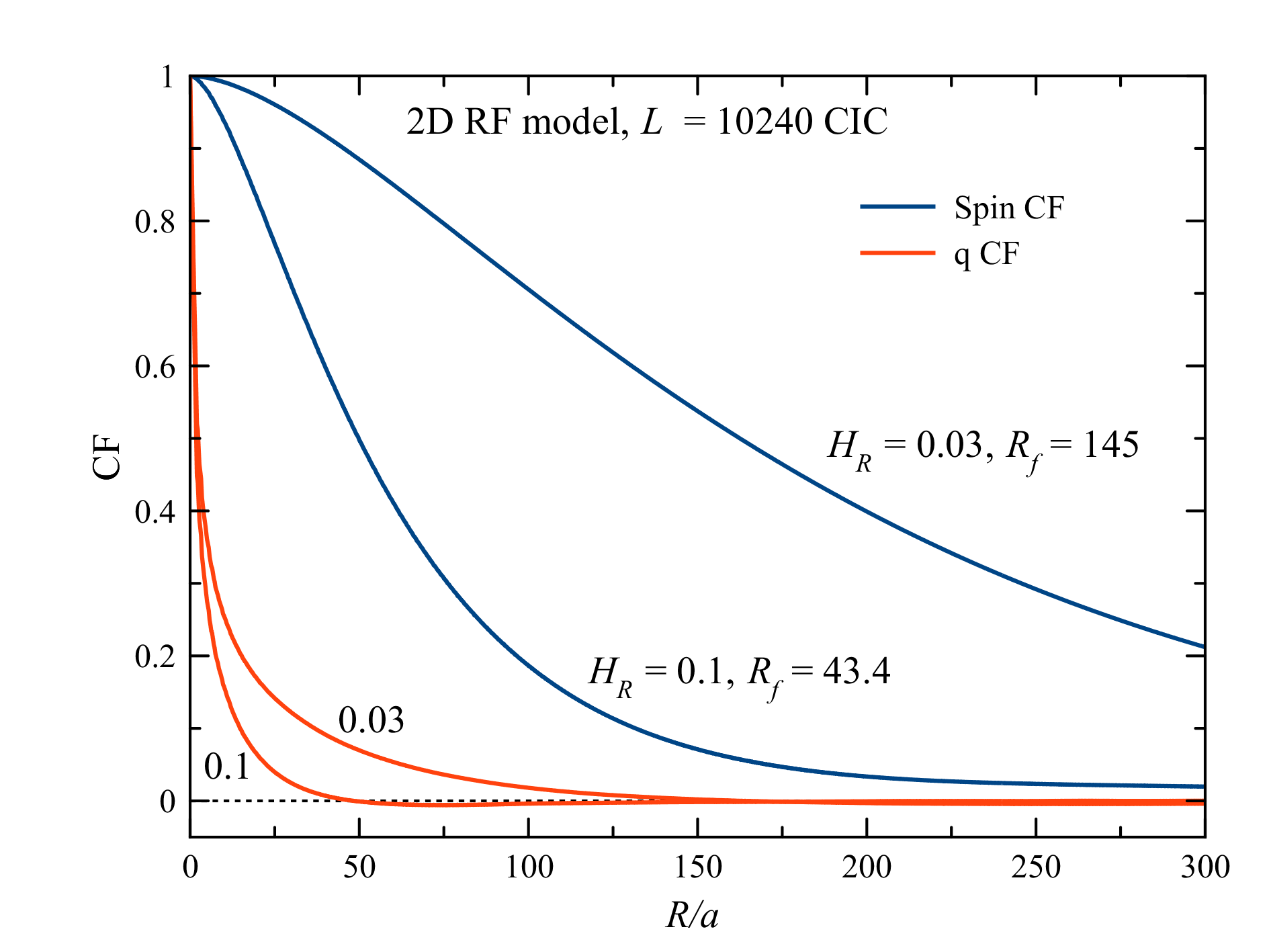}
\includegraphics[width=90mm]{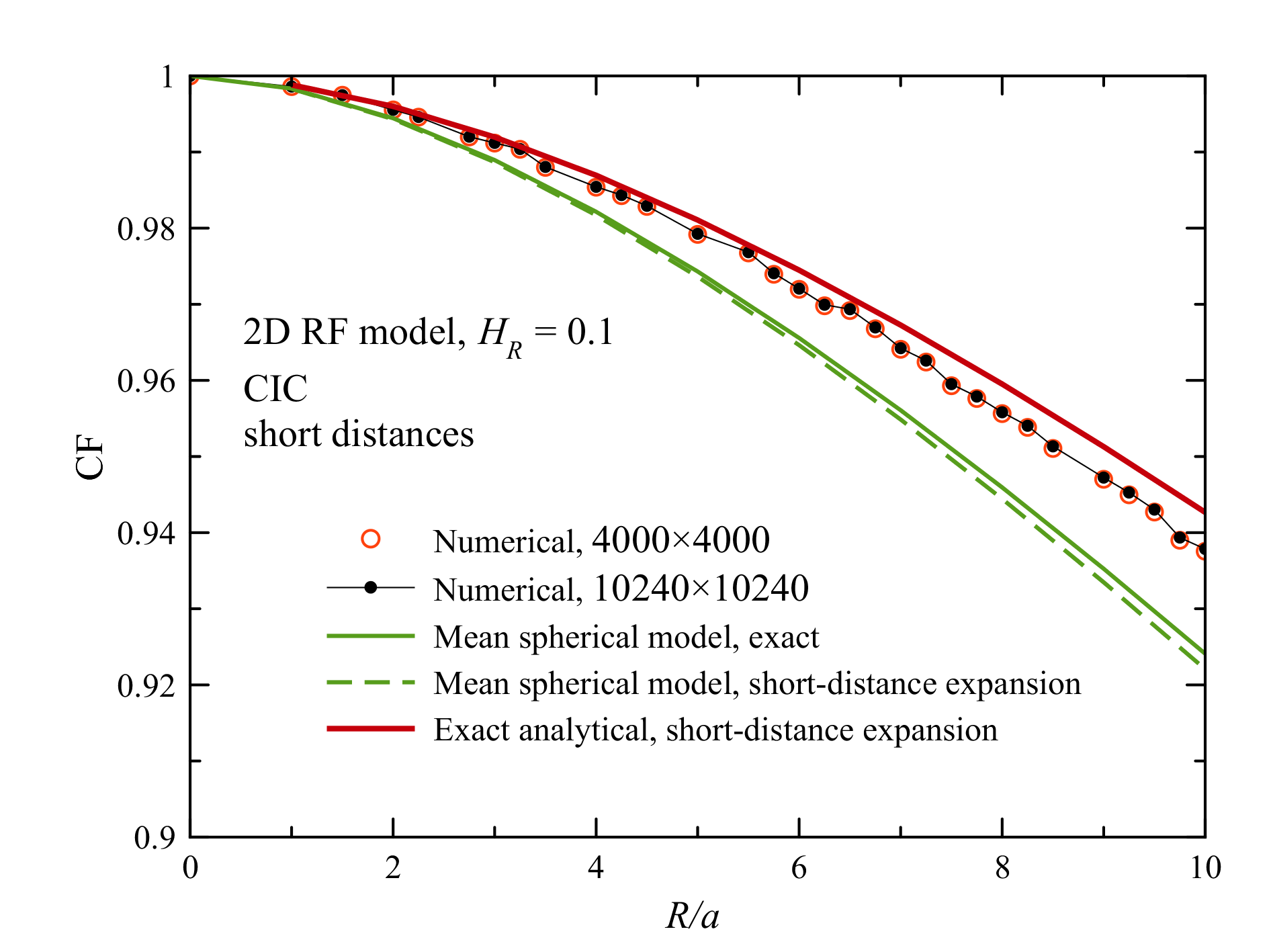}
\caption{Color online: Upper panel: Spin-spin (blue) and TCD (red) CFs for two values of the random field, computed with the CIC on a square lattice $10240 \times 10240$. Lower panel: Short-range behavior of the spin-spin CF.}
\label{r-CFs}
\end{figure}

Numerically obtained real-space spin-spin and TCD CFs vs $R = |{\bf r}_1 - {\bf r}_2|$ are shown in the upper panel of Fig. \ref{r-CFs} for two values of $H_R$. The TCD CF drops much faster than the spin-spin CF. A more careful analysis (see below) shows that it changes sign at $R \approx R_f$ and then oscillates on increasing R.  Values of the ferromagnetic correlation length, $R_f$, that appear in the upper panel of Fig. \ref{r-CFs}, are taken from the theoretical formula derived below. They provide a good fit of the short-range behavior of the CF shown in the lower panel of Fig. \ref{r-CFs}. 

While the system is highly nonlinear it is instructive to compare the numerical results with the analytical results for the spin-spin CF that can be obtained if one ignores topological defects.  One such possibility is presented by the 2D mean-spherical model in which the spin field ${\bf S}({\bf r})$ is allowed to have an arbitrary length while satisfying an integral condition $\langle {\bf S}^2\rangle = S_0^2$. It is statistically equivalent to the spin field with an infinite number of spin components \cite{Stanley}, in which case ($n > d + 1$) the topological objects are absent \cite{PGC-PRL2014}. Adding the term $ - \Lambda \int d^2 r \,{\bf S}^2$ with the Lagrange multiplier $\Lambda \equiv -\alpha k_f /2$ to the Hamiltonian one obtains the following equation for the spin field: $({\bm \nabla}^2 - k_f^2){\bf S} = -{\bf H}_R/\alpha$. Its solution is ${\bf S}({\bf r}) = -\alpha^{-1} \int d^2r'G({\bf r} - {\bf r}'){\bf H}_R({\bf r}')$, where  $G({\bf r})$ is the Green function of the differential equation for ${\bf S}$, having a Fourier transform $G({\bf k}) =-{1}/({k^2 + k_f^2})$. Writing for the RF $\langle{H}_{Ri}({\bf r}'){H}_{Rj}({\bf r}'')\rangle =\frac{1}{3}
H_R^2a^2\delta_{ij}\delta(|{\bf r}' - {\bf r}''|)$, one gets for ${\bf s} = {\bf S}/S_0$
\begin{equation}\label{MS-CF}
\langle{\bf s}({\bf r}_1)\cdot{\bf s}({\bf r}_2)\rangle = (k_f R)K_1(k_f R)  , \quad    k_f a = \frac{H_R}{2\sqrt{\pi} Js},
\end{equation}
where $k_f$ was obtained from the condition $\langle {\bf s}^2\rangle = 1$. Here $K_1(x)$ is a modified Bessel function having asymptotes $K_1(x) \rightarrow 1/x$ at $x \rightarrow 0$ and $K_1(x) \rightarrow (\pi/2x)^{1/2}\exp(-x)$ at $x \gg 1$. The more accurate expansion at short distances, $R \ll R_f \equiv 1/k_f$, is
\begin{equation}\label{short}
\langle{\bf s}({\bf r}_1)\cdot{\bf s}({\bf r}_2)\rangle \rightarrow 1 - [R/(2R_f)]^2\ln(2R_f/R).
\end{equation}
In that limit, however, one can develop a more rigorous approach that agrees with numerics quantitatively. The exact equation for ${\bf s}({\bf r})$ is
\begin{equation}\label{equation-S}
 \alpha {\bm \nabla}^2 {\bf s} - {\alpha}{\bf s}({\bf s}\cdot {\bm \nabla}^2 {\bf s}) + {\bf H}_R - {\bf s}({\bf s}\cdot {\bf H}_R) = 0.
\end{equation}
At short distances, starting with ${\bf s} = {\bf s}_0$ at a certain point and writing ${\bf s} = {\bf s}_0 + \delta {\bf s}$ in the vicinity of that point, it is easy to see that due to the nonsingular nature of skyrmions a weak rotation of ${\bf s}$ always provides a $\delta {\bf s}$ smallness of ${\bf s}\cdot{\bm  \nabla}^2{\bf s} = -({\bm \nabla} \delta {\bf s})^2$ as compared to ${\bm \nabla}^2{\bf s} = {\bm \nabla}^2 \delta {\bf s}$. This allows one to neglect the second term in Eq.\ (\ref{equation-S}), reducing it to $\alpha \nabla^2 {\bf s} = - {\bf H} + {\bf s}({\bf s}\cdot {\bf H}_R)$ that can be writen in the integral form ${\bf s}({\bf r}) = -{\alpha}^{-1} \int d^2r'G({\bf r} - {\bf r}'){\bf g}({\bf r}')$, with the Fourier transform of $G({\bf r})$ being $-1/k^2$ and ${\bf g} =  {\bf H}_R - {\bf s}({\bf s}\cdot {\bf H}_R)$. At this point the CF at short distances can be computed for any $n$-component spin. Noticing that  $\langle{\bf g}({\bf r}')\cdot {\bf g}({\bf r}'')\rangle =  {H_R^2}(1 -1/n)a^2\delta({\bf r}' - {\bf r}'')$, one obtains Eq.\ (\ref{short}) but with a different $R_f$,
\begin{equation}\label{R-n}
\frac{R_f}{a} = 2\left(\frac{\pi}{1-1/n}\right)^{1/2} \frac{Js}{H_R}.
\end{equation}
As expected, at $ n = \infty$ it yields $R_f$ of the 2D mean-spherical model while at $n = 3$ a slightly different result follows: $R_f = (6\pi)^{1/2}(Js/H_R)$. It gives $R_f/a \approx 43.4$ for $H_R = 0.1$ and $R_f \approx 145 $ for $H_R = 0.003$, which agrees remarkably with the numerical fit at short distances. At large distances the spin-spin CF exhibits some kind of exponential decay with $R_f \propto 1/H_R$, although its exact analytical form remains unknown. For, e.g., $H_R = 0.1$ the CF decreases in half at $R/a = 52$.

\begin{figure}
\includegraphics[width=90mm]{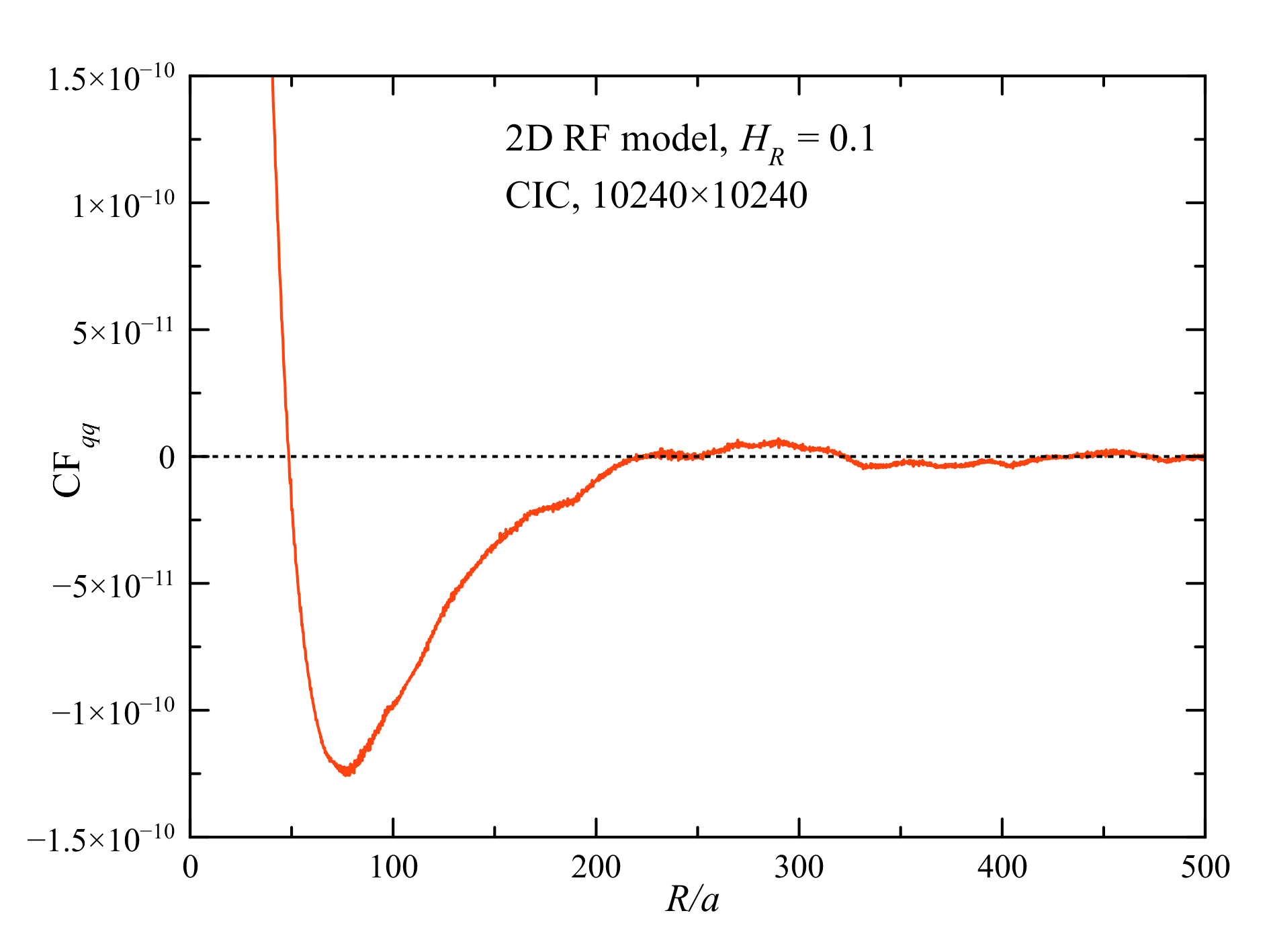}
\includegraphics[width=95mm]{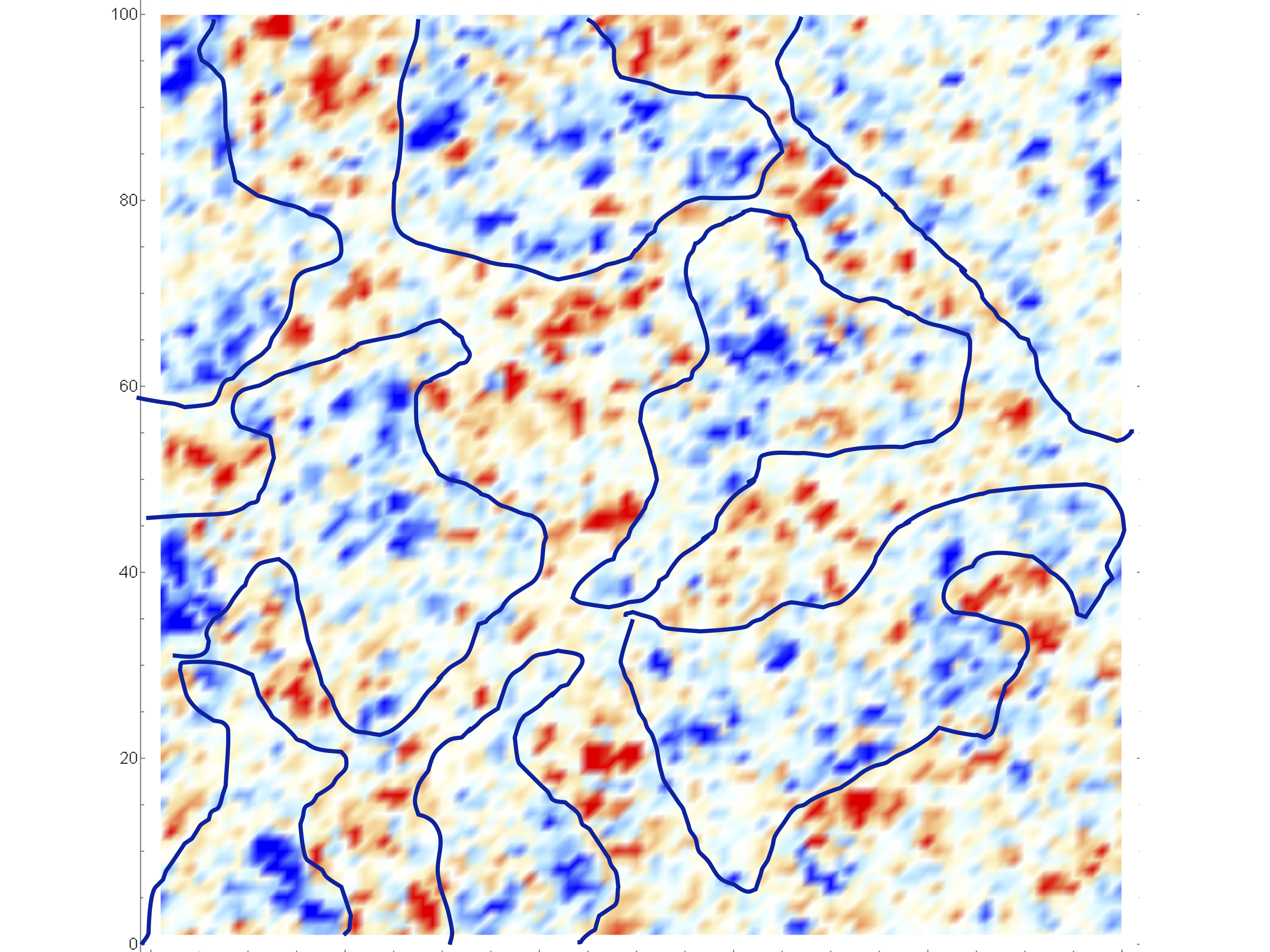}
\caption{Color online: Upper panel: Oscillations of the TCD CF at large distances. Lower panel: TCD plot. Red/blue and the density of the color show the sign and the magnitude of the TCD correspondingly. Solid lines are guidance for the eye to see the grainy structure of the topological charge associated with IM domains.}
\label{TCD}
\end{figure}
We now focus our attention on the TCD CF. Its behavior at large distances is shown in the upper panel of Fig. \ref{TCD}. Unlike the spin-spin CF the TCD CF oscillates at large distances with decreasing amplitude. The parameter $R_f \propto 1/H_R$ defines the period of the oscillations. For, e.g., $H_R = 0.1$ the first zero occurs at $R/a \approx 46$ which is pretty close to $R_f$ obtained for the spin-spin CF. Since the latter provides the average size of the region where the spins are ferromagnetically correlated, it shows some kind of the oscillating topological order associated with the IM domains: Domains with a positive topological charge are surrounded  by domains with a negative topological charge. The latter is illustrated by the plot of the TCD shown in the lower panel of Fig. \ref{TCD}.  These results hint that the correlated regions could be formed by coupled skyrmions and antiskyrmions deformed by their interaction and by the RF energy landscape. 

\begin{figure}
\includegraphics[width=90mm]{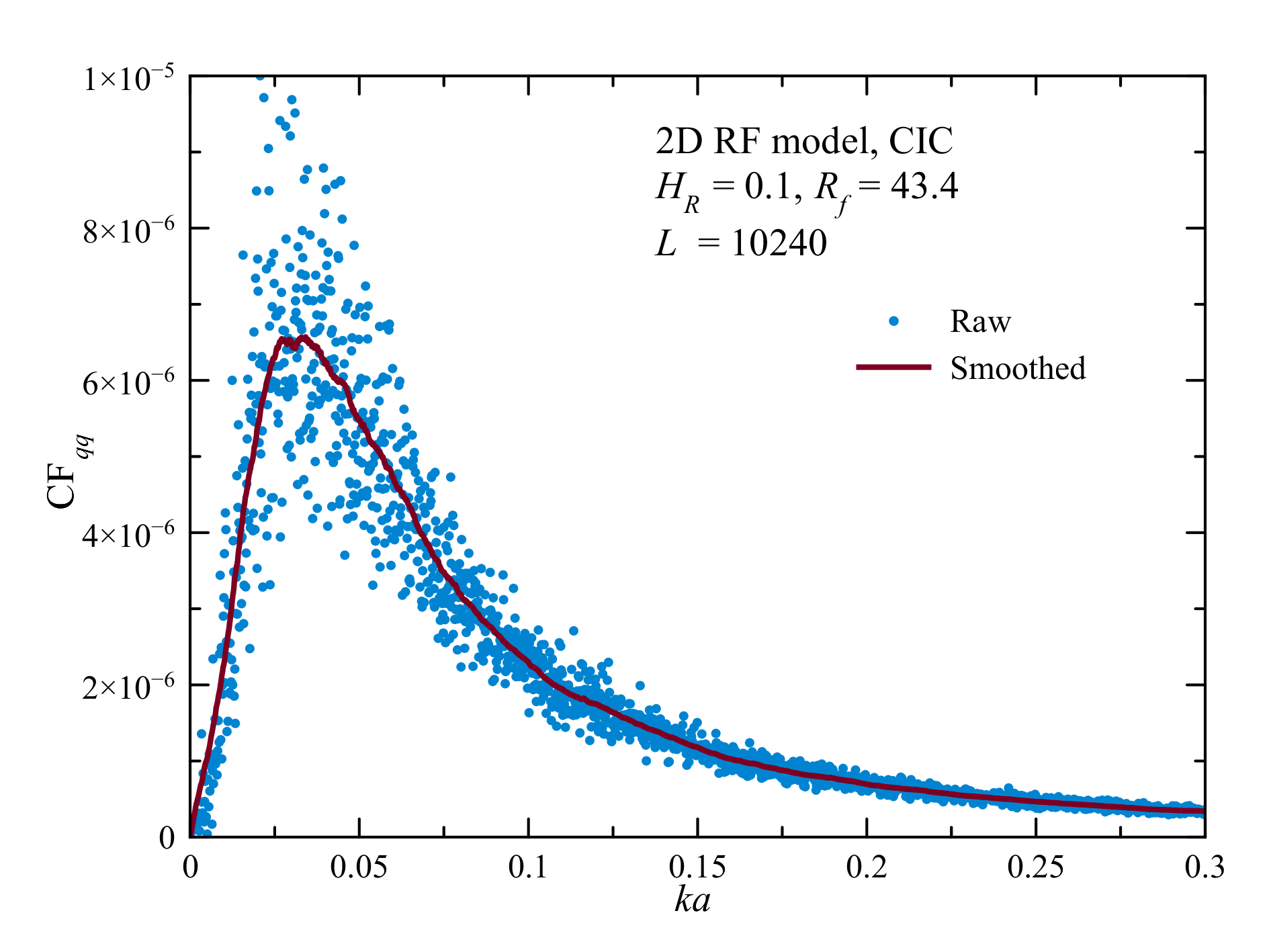}
\includegraphics[width=90mm]{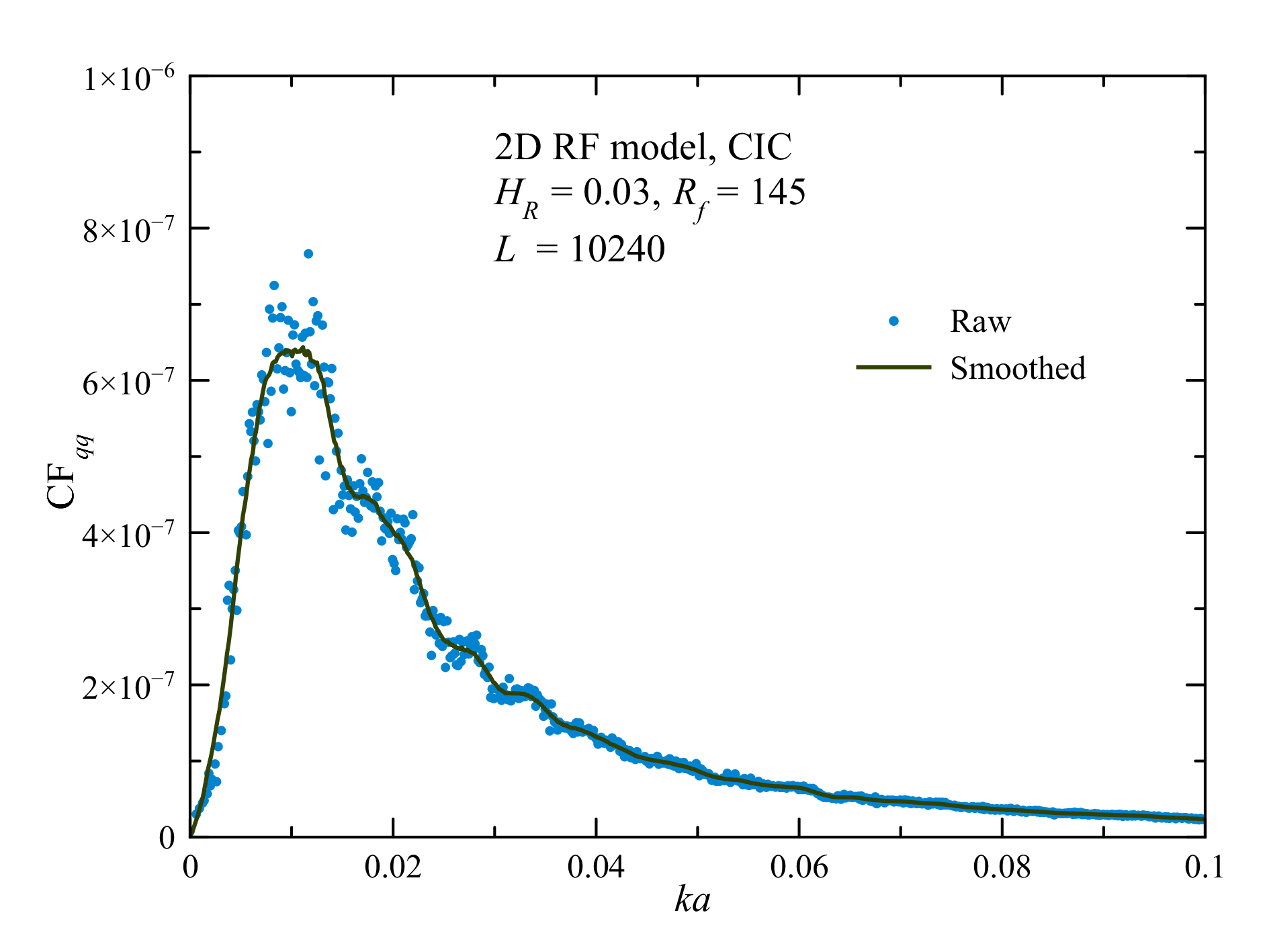}
\caption{Color online: Fourier transform of the TCD CF (raw and smoothed) for $H_R = 0.1$ (upper panel) and $H_R = 0.03$ (lower panel).}
\label{q-TCD}
\end{figure}
Further evidence of the topological order associated with the IM domains comes from the Fourier transform of the TCD CF shown in Fig. \ref{q-TCD}. It exhibits a maximum at $kR_f = 1$, thus confirming the oscillating structure of the TCD. To relate the observed oscillations of the TCD to skyrmions one can estimate the absolute value of the topological charge of the correlated area as $Q_{\rm CA} = (2R_f/L)^2 \int d^2 r |q|$ where $L\times L$ is the total area of the 2D system. At $H_R = 0.1$ this gives $Q_{\rm CA} = 0.948$ for $R_f/a = 43.4$ (the short-range result for the spin-spin CF) and $Q_{\rm CA} = 1.065$ for $R_f/a  = 46$ (the first zero of the TCD CF). Both values of $Q_{\rm CA}$ are pretty close to the skyrmion charge $Q = 1$. The exchange energy of the correlated area coincides with the ground state energy of the skyrmion, $4\pi J$, up to a factor of order unity. 

In conclusion, we have provided evidence that a static random field in a 2D exchange model transforms the ordered state into a  skyrmon-antiskyrmion glass.  Experimental detection of the topological order requires accurate mapping of the directions of spins in large areas. It could be worth the effort because it would help to solve the fundamental problem of the nature of the glass state in systems with nontrivial topology. Recent experiments on skyrmions in disordered films with ferromagnetic exchange \cite{2D-amorphous,Montoya-PRB2017} make the first step in that direction. 

This work has been supported by the grant No. DE-FG02-93ER45487 funded by the U.S. Department of Energy, Office of Science.

\end{document}